\documentclass[preprint,showpacs,preprintnumbers,amsmath,amssymb, superscriptaddress,longbibliography,nofootinbib]{revtex4-1}

\usepackage{textcomp}
\usepackage{makeidx}
\usepackage{amsmath}
\usepackage{subfig}
\usepackage{amssymb}
\usepackage{hyperref}
\usepackage{graphicx}
\usepackage{epstopdf}
\usepackage{color}
\usepackage{soul}

\begin{document}

\title{A new class of regular Black Holes in Einstein Gauss-Bonnet gravity with localized sources of matter}

\author{Milko Estrada}
\email{milko.estrada@gmail.com}
\affiliation{Facultad de Ingenier\'ia, Ciencia y Tecnolog\'ia, Universidad Bernardo O'Higgins, Av. Viel 1497, Santiago 8370993, Chile}
\author{Rodrigo Aros}
\email{raros@unab.cl}
\affiliation{Departamento de Ciencias Fisicas, Universidad Andres Bello, Av. Republica 252, Santiago 8370134, Chile}

\date{\today}

\begin{abstract}
We provide a new regular black hole solution (RBH) in Einstein Gauss-Bonnet (EGB) gravity with localized sources of matter in the energy--momentum tensor. We determine the necessary constraints in order for the solution to be regular. Although we use a specific form for the energy density as a test of proof, these constraints could serve as a recipe for constructing several new RBH solutions in EGB gravity with localized sources. Due to that the usual first law of thermodynamics is not valid for RBH, so we rewrite the first law for EGB, which leads to correct values of entropy and volume. The size of the extremal black hole, whose temperature vanishes, becomes smaller for larger dimensions, whose radius could be of the order of the Planck units, thus the evaporation would stop once the horizon radius contracts up to a value close to the Planck length, which could be related with the apparition of quantum effects. Furthermore, the presence of matter fields in the energy--momentum tensor induces two phase transitions, where there are two regions of stability. This differs from the vacuum EGB solution, where the specific heat is always negative without phase transition as occurs in Schwarzschild black hole.
\end{abstract}

\maketitle

\section{Introduction}
Several branches of theoretical physics have predicted the existence of extra dimensions. However, nowadays, scientists have observed only four-dimensional gravitational scenarios. So, all theories that consider the existence of extra dimensions are expected to coincide with General Relativity in four dimensions, which is the valid four-dimensional gravitational theory today. Between these theories, we find the Lovelock gravity \cite{Lovelock:1971yv}. The Lagrangian of Lovelock gravity considers the inclusion of higher curvature terms as corrections to the Einstein-Hilbert action
\begin{equation}\label{LovelockLagrangian}
\displaystyle L  = \frac{1}{16 \pi G}   \sum_{k=0}^{[d/2]} \alpha_k L_k,
\end{equation}
with
\begin{equation}
    L_k=\frac{1}{2^{k}}\,\sqrt{-g}\ \delta_{\nu_{1}...\nu_{2k}}^{\mu_{1}...\mu_{2k}}\ R_{\mu_{1}\mu_{2}}^{\nu_{1}\nu_{2}}\cdots R_{\mu_{2k-1}\mu_{2k}}^{\nu_{2k-1}\nu_{2k}} ,
\end{equation}
where $d$ is the number of dimensions, $G$ is the Newton constant, $R^{\alpha \beta}_{\hspace{2ex}\mu\nu}$ is the Riemann tensor, $L_k$ is the Euler Density of order $k$ in the Riemann tensor, $\delta^{\mu_1\ldots \mu_k}_{\nu_1 \ldots  \nu_k}$ is the  $k$-anti-symmetric generalized Kronecker delta and $\{\alpha_k\}$ is an arbitrary set of coupling constants. The first three terms of the Lagrangian are proportional to the cosmological constant, to the Ricci scalar, and to the Einstein Gauss-Bonnet term. For the Einstein Gauss-Bonnet (EGB) case, the sum in equation \eqref{LovelockLagrangian} is computed up to $k=2$. It is worth mentioning that EGB gravity has called attention in the last years for theories of inflation and contrasted with the GW170817 results \cite{Oikonomou:2021kql,Odintsov:2020zkl} .

In even dimensions $d=2k$, the Euler density of order $k=d/2$ is a  topological invariant and therefore does not contribute to the equations of motion. For both even and odd dimensions the term $k<d/2$ contributes to the equations of motion. For $k>d/2$ the Euler density vanishes. In this respect, Lovelock's equations of motion are equivalent to those of General Relativity for spacetimes with three and four dimensions. Furthermore, Lovelock's theories respect the basic principles of General Relativity, for instance, its equations of motion are of second order. It is worth mentioning that in a different approach recently in \cite{Glavan:2019inb} was proposed, by a rescaling of the coupling constant $\alpha_2$, that the Gauss-Bonnet term could give rise to a contribution to the field equations in four dimensions. This, in turn, gave rise to several new results. See for example references \cite{Lu:2020iav,Fernandes:2020nbq,Fernandes:2021dsb,Ovgun:2021ttv,Aoki:2020lig}.

On the other hand, the recent detection of gravitational waves through the collision of two rotating black holes \cite{LIGOScientific:2016aoc} has positioned these latter as one of the most exciting and intriguing objects in gravitation. It is well known the fact that the black hole solutions have a central singularity, which represents a point where both the metric as the invariants of curvature diverge, and so, in this point, the laws of physics are not valid. The first model to address this problem was formulated by Bardeen in reference \cite{bardeen}. In this reference, from the classical point of view, the singularity formation is avoided by changing the mass parameter in the Schwarzschild solution by a radial function, such that near the origin the solution behaves as a de-Sitter spacetime. The models without the presence of singularities are known as {\it regular black holes} (RBH). After this, were showed RBH solutions based on the inclusion of nonlinear electrodynamics sources in the energy-momentum tensor. The first of these latter models was demonstrated in reference \cite{Ayon-Beato:2000mjt}, where it is shown that the Bardeen solution can be interpreted as a magnetic solution to Einstein equations coupled to nonlinear electrodynamics. In Einstein Gauss Bonnet coupled to nonlinear electrodynamics for $d=5$ was found a generalization of the Bardeen solution in reference \cite{Singh:2019wpu}. We can see a review of the inclusion of nonlinear sources in gravitation in reference \cite{Sorokin:2021tge}. We can see some examples of recent models of RBH with nonlinear electrodynamics sources in references \cite{Mkrtchyan:2022ulc,Malafarina:2022oka,Sajadi:2017glu,Sajadi:2019hzo,Hendi:2020knv,Sajadi:2021ilt}.

Another type of RBH consists on the inclusion of localized sources of matter in the energy-momentum tensor, where the energy density has local values for different values of the radial coordinate. In these models, the energy density tends to zero at infinity and so these models have a well-defined asymptotic structure. In 2019, in reference \cite{Aros:2019quj}, for higher dimensional scenarios in Pure Lovelock (PL) \cite{Cai:2006pq} and Lovelock with $n-$fold degenerated ground state (LnF) \cite{Crisostomo:2000bb} theories, were established a list of constraints in order that the solution be regular and was proposed a $d-$dimensional model of energy density such that the solution is free of singularities. This energy density model for four-dimensional scenarios can be rewritten as the Hayward metric \cite{DeLorenzo:2014pta}. Furthermore, this model of energy density was studied for the $d=(2+1)$ case in Einstein Hilbert theory in reference \cite{Estrada:2020tbz} and in reference \cite{Maluf:2022jjc} was showed that the nature of this type of energy density is incompatible with the behavior of the nonlinear electrodynamics sources. A recent generalization of this model of energy density in reference \cite{Hendi:2022opt}.

However, both PL and LnF theories correspond to particular cases of Lovelock gravity with a particular ground state structure and so with a particular structure of equations of motion. In this regard, Einstein Gauss-Bonnet (EGB) has a different structure of equations of motion \cite{Boulware:1985wk} with respect to the cases before mentioned and furthermore its form of entropy also differs \cite{Myers:1988ze} in the vacuum case with respect to other theories of gravity. So, it is of physical interest to test the necessary physical requirements in order to obtain RBH solutions in EGB gravity with localized sources of matter. Furthermore, it is of physical interest to test how varies the physical and thermodynamic properties of the RBH solutions in EGB gravity with localized sources of matter with respect to RBH solutions in other scenarios. 

On the other hand, the usual version of the first law of thermodynamics, namely $dM=TdS+PdV$ is not valid for RBH due to the presence of matter fields in the energy--momentum tensor \cite{Ma:2014qma}. This is because the usual form of the first law leads to incorrect values of entropy and volume. So, it is also of physical interest to test if it is possible to rewrite the first law of thermodynamics for RBH in EGB gravity in order to obtain correct values of entropy and volume.

In this article, we will test the necessary constraints in order to obtain an RBH solution in EGB gravity with the presence of localized sources of gravity in the energy--momentum tensor. So, we will present a new class of RBH solutions in EGB gravity. Furthermore, we will propose a version of the first law of thermodynamics for RBH in EGB gravity and we will analyze the thermodynamics behavior of the solution.

\section{The equations of motion}
We study the following spherically symmetric spacetime.
\begin{equation}
ds^2 = -f(r) dt^2 + f(r)^{-1} dr^2 + r^2 d\Omega_{D-2}
\end{equation}
where $d\Omega_{D-2}$ corresponds to the transversal section of a $D-2$ sphere. The equation of motion are \cite{Hull:2021bry}:
\begin{equation} \label{eqmovt}
\displaystyle  \delta^i_j  \sum_{n=0}^{n=K} \left ( b_n \left ( r^{d-2n-1} A_n \left ( - \frac{f(r)}{r^2} \right ) \right ) \right) = \frac{16 \pi G}{(d-2)} \bar{m}(r)
\end{equation}
where
\begin{equation} \label{funciondemasa0}
    \bar{m}(r)= - \int r^{d-2} T^i_j dr
\end{equation}
and
\begin{equation}
\displaystyle - \frac{\delta^\alpha_\beta}{2 r^{d-3}} \frac{d^2}{dr^2}  \sum_{n=0}^{n=K} \left ( b_n \left ( r^{d-2n-1} A_n \left ( - \frac{f(r)}{r^2} \right ) \right ) \right) = 8 \pi G T^\alpha_\beta
\end{equation}
where the index $i,j$ take values $i,j=0,1$ and $\alpha, \beta=2...D-1$, and where $A_n \left ( - \frac{f(r)}{r^2} \right )$ is computed using the following polynomial
\begin{equation}
    A_n \left ( - \frac{f(r)}{r^2} \right ) =  \sum_{k=n}^{K} \alpha_k \binom{n}{k} \left ( - \frac{f(r)}{r^2} \right )^{k-n}
\end{equation}

It is worth mentioning that, for our case, where the transversal section corresponds to a $(D-2)$ sphere, all the constant $b_n$ are equal to $b_n=1$.

The energy-momentum tensor (EM) has the form
\begin{equation}\label{TensorEM}
    T^{\alpha}_{\hspace{1ex}\beta} =  \textrm{diag}(-\rho, p_r, p_\theta, p_\theta, ...).
\end{equation}
where it is easy to check from equation \eqref{eqmovt} that, from this form of the EM tensor $T_0^0=T_1^1 \to -\rho=p_r$. The components $(0,0)$ and $(1,1)$ of the equations of motion are similar. So, the equation \eqref{funciondemasa0} can be rewritten as:
\begin{equation} \label{funciondemasa1}
    \bar{m}(r)= \int  r^{d-2} \rho (r) dr
\end{equation}

It is worth mentioning that in the space of parameters, such that $f(r_h,M)=0$, we will consider that $\bar{m}=\bar{m}(r_h,M)$, where $r_h$ and $M$ will be related with the horizon radius and the mass.

\section{The EGB solution with localized sources of matter} \label{SolucionGenerica}

Setting $K=2$, and for simplicity, we define:

\begin{equation} \label{funciondemasa2}
    m(r)=\frac{8 \pi \bar{m}(r)}{(d-2)}
\end{equation}

Evaluating the equation \eqref{eqmovt} we obtain the following polynomial dependent of $f = f(r)$

\begin{equation}
    \alpha_2 f^2 + (-\alpha_1 r^2 - 2 \alpha_2) f + \alpha_0 r^4 + \alpha_1 r^2 + \alpha_2 = \frac{2 G m(r)}{r^{d-5}}
\end{equation}
where, as was mentioned in the introduction, $d \ge 5$, due that for $n=d/2=2$ with $d=4$ the lagrangian of Gauss Bonnet $L_{n=2}$ is a topological invariant. The solution is

\begin{equation} \label{Solucionpm}
    f(r)=1 + r^2 \frac{\alpha_1}{2\alpha_2} \pm \frac{\sqrt{r^4 (\alpha_1^2-4 \alpha_0 \alpha_2) +   8 \alpha_2 G \dfrac{m(r)}{r^{d-5}}                   }}{2\alpha_2}
\end{equation}

It is worth noting that we choose the following convention for the units system: $[G]=\ell^{d-2}$, $[\alpha_k]=\ell^{2k-2} \to [\alpha_0]=\ell^{-2}, [\alpha_1]=\ell^{0}, [\alpha_2]=\ell^{2}$, where $\ell$ represents units of length. So, it is necessary that $[m(r)]=\ell^{-1}$.

It is worth noting that the solution \eqref{Solucionpm} has no small limit when $\alpha_2 \to 0$. So, it is not possible to recover the General Relativity solution in this case. Thus, in this work, we will study the branch with $-$ sign
\begin{equation} \label{Solucion}
    f(r)=1 + r^2 \frac{\alpha_1}{2\alpha_2} - \frac{\sqrt{r^4 (\alpha_1^2-4 \alpha_0 \alpha_2) +   8 \alpha_2 G \dfrac{m(r)}{r^{d-5}}}}{2\alpha_2}
\end{equation}
On the other hand, the limit $\alpha_2 \to 0$ is:
\begin{equation} \label{asintota}
    f |_{\alpha_2 \approx 0} \approx 1 + \frac{\alpha_0}{\alpha_1} r^2-\frac{G}{\alpha_1} \frac{2m(r)}{r^{d-3}}
\end{equation}
which coincides with the solution corresponding to General Relativity.

\subsection{A list of criteria for the solution} \label{ListaCriterios}
Due to the presence of the energy density, {\it i.e.} of the mass function $m(r)$ in the solution, the study of the conditions for the solution in order to be regular is not trivial. Furthermore, both the value of the coupling constant $\alpha_0,\alpha_1$ and $\alpha_2$ as the form of the solution and the equations of motion differs from the RBH solution in the special cases Pure Lovelock and Lovelock with Unique Vacuum \cite{Aros:2019quj}. So, it is necessary to analyze the features of the solution and determine the necessary constraints in order for the solution to be regular: 

\begin{itemize}
    \item In order that the expression \eqref{funciondemasa1} be consistent, it is necessary that the energy density $\rho(r)$ be free of singularities everywhere.
    \item In order that the limit $\alpha_2 \approx 0$ has a correct asymptotic behavior it is necessary that
    \begin{equation} \label{CondicionM}
        \displaystyle \lim_{r\to \infty} m(r)= \mbox{Constant}=M>0.
    \end{equation}
This implies that, at infinity:
    \begin{equation} \label{Condicionrho}
        \displaystyle \lim_{r\to \infty} \rho(r)= 0
    \end{equation}
So, at infinity the equation \eqref{asintota} behaves like the Schwarzschild--Tangherlini solution with the presence of a cosmological constant.

\item In order that the limit $\alpha_2 \approx 0$, equation \eqref{asintota} recovers the attractive Newtonian potential at infinity it is necessary that: 
\begin{equation} \label{Condicionalpha1}
    \alpha_1 >0
\end{equation}
Furthermore, in order that equation \eqref{asintota} to have an Anti-de--Sitter asymptotic behavior, and thus, it is avoided the presence of a cosmological horizon,  it is necessary that
\begin{equation} \label{Condicionalpha0}
    \alpha_0 >0
\end{equation}

\item In order that the metric potential \eqref{Solucion} be well defined for all values of $r$, {\it i.e} the interior of the square root be always positive and free of singularities, we impose that
\begin{align}
    \alpha_1^2-4 \alpha_0 \alpha_2 &>0 \label{CondicionTerminos}\\
    \alpha_2 &>0 \label{Condicionalpha2} \\
    m(r) &>0 \label{Condicionm} \\
    m(r) |_{r \approx 0} & \approx M r^N \mbox{\, , with $N \ge d-5$ and $M>0$ a constant} \label{Condicionmmetrica}
\end{align}
where $m(r)$ is an increasing function such that the equation \eqref{CondicionM} is satisfied.
\item It is worth mentioning that conditions \eqref{CondicionM}, \eqref{Condicionrho}, \eqref{Condicionalpha1}, \eqref{CondicionTerminos} and \eqref{Condicionalpha2} ensure that the solution \eqref{Solucion} has a correct behavior at the asymptote.

Following the conditions \eqref{CondicionM} and \eqref{Condicionrho}, it is possible to define an effective cosmological constant at infinity such that the solution \eqref{Solucion} behaves as
\begin{equation} \label{Solucioninfinito}
    f(r)|_{r \to \infty} \to 1- r^2 \frac{1}{2\alpha_2} \left (-\alpha_1 + \sqrt{ (\alpha_1^2-4 \alpha_0 \alpha_2)} \right ) =1 - r^2\Lambda_{eff}^{(r \to \infty)}
\end{equation}
where, it is easy to check that $\Lambda_{eff}^{(r \to \infty)}<0$, and thus, the spacetime has an AdS asymptote.
\end{itemize}

Given the conditions \eqref{Condicionalpha1}, \eqref{Condicionalpha0}  \eqref{CondicionTerminos} and \eqref{Condicionalpha2}, it is easy to check that the effective cosmological constant at infinite is negative, so, the solution has an AdS asymptote. Thus, it is possible to determine the conserved charge of the solution up to infinity \cite{Mora:2004rx}.

The Kretschmann scalar is:
\begin{equation}
K =\left (f''(r) \right)^2 + \frac{2(d-2)}{r^2} \left (f'(r) \right)^2 + \frac{2(d-2)(d-3)}{r^4} \left (1-f(r) \right)^2
\end{equation}

\begin{itemize}
    \item In order to have a solution free of singularities near the origin, we choose that the solutions behave as an (A)dS or flat spacetime near the origin, {\it i.e} $f|_{r \approx 0} \approx 1 \pm C r^2$ or $f|_{r \approx 0} \approx 1$. In future works could be testing if it is possible to compute a solution whose behavior near the origin is $f|_{r \approx 0} \approx 1 \pm C r^N$, whit $N>2$.

    For this, we can check from equation \eqref{Solucion} that the behavior of the mass function must be:
    \begin{equation} \label{funcionmasa0}
    m(r)|_{r \approx 0}  \approx Mr^{d-1}
    \end{equation}
    This is consistent with condition \eqref{Condicionmmetrica}. Below we will discuss the core structure.
 \item Furthermore we need that both $f(r)$ as its first and second derivatives be free of singularities for $r\ge 0$. This is ensures with the conditions \eqref{CondicionTerminos}, \eqref{Condicionalpha2}, \eqref{Condicionm} and \eqref{funcionmasa0}.
\end{itemize}

It is worth mentioning that, these constraints could serve as a recipe for constructing several new RBH solutions in EGB gravity with localized sources of matter in the energy--momentum tensor. Below we will show an example as a test of proof. 

\subsection{Core structure}
From equations \eqref{Solucion} and \eqref{funcionmasa0}, the behavior of the solution near the origin is:
\begin{equation} \label{Solucionnear0}
    f(r)|_{r \approx 0} \approx 1- r^2 \frac{1}{2\alpha_2} \left (-\alpha_1 + \sqrt{ (\alpha_1^2-4 \alpha_0 \alpha_2) +   8 M \alpha_2 G } \right ) =1 - r^2\Lambda_{eff}^{(r \approx 0)}
\end{equation}
where the conditions \eqref{CondicionM}, \eqref{CondicionTerminos} and \eqref{Condicionalpha2} ensure that the interior of the square root be positive and where $\Lambda_{eff}^{(r \approx 0)}$ represents an effective cosmological constant near the origin. Thus, the core of the solution could have:
\begin{itemize}
    \item A de-Sitter structure for $\sqrt{ (\alpha_1^2-4 \alpha_0 \alpha_2) +   8 M \alpha_2 G }>\alpha_1$. It is worth mentioning that, in the hypothetical case where the energy density is of order of the Planck units near the origin,  the positive effective cosmological constant could be viewed as a repulsive force due to quantum effects \cite{DeLorenzo:2014pta}.
     \item An Anti de-Sitter structure for $\sqrt{ (\alpha_1^2-4 \alpha_0 \alpha_2) +   8 M \alpha_2 G }<\alpha_1$.
     \item An flat structure for $\sqrt{ (\alpha_1^2-4 \alpha_0 \alpha_2) +   8 M \alpha_2 G }=\alpha_1$.
\end{itemize}
It is worth mentioning that the search for an explanation connecting both AdS as flat cores with the nature of a Planckian energy density near the origin is an open question of deep physical interest \cite{Christiansen:2022ebo} , which could be studied in future work.

\subsection{Thermodynamics Features}
\subsubsection{ \bf A brief comment about the first law of thermodynamics}

It is well known the fact that the usual form of the first law of thermodynamics, namely $dM=TdS-PdV$ must be modified for regular black holes due to the presence of fields of matter in the energy--momentum tensor \cite{Ma:2014qma}. This is because this latter leads to incorrect values of entropy and thermodynamics volume (as for example in GR the value of entropy does not  follow the area's law). With respect to this, in the literature have had efforts in order to address this problem. In reference \cite{Ma:2014qma} was included a correction factor, which corresponds to an integration of the radial coordinate up to infinity. In reference \cite{Maluf:2022jjc} was included in the first law an extra thermodynamic potential $dX$, and the thermodynamics pressure was identified with the radial pressure of the EM tensor, in order to recover the correct values of entropy and volume. In \cite{Estrada:2020tbz}, for Einstein Hilbert gravity in $(2+1)$ dimensions, was defined a local definition of the variation of energy at the horizon by imposing constraints on the evolution along the space parameters. In this work, following the idea of \cite{Estrada:2020tbz}, but now for a $d-$dimensional spacetime, and for Einstein Gauss-Bonnet gravity, we will test if it is possible to find a version for the first law for regular black holes. Specifically, we will use the conditions $f(r_h,M,\alpha_0)=0$ and $\delta f(r_h,M,\alpha_0)=0$ :
\begin{equation}
    0 = \frac{\partial f}{\partial r_+} dr_+ + \frac{\partial f}{\partial M} dM + \frac{\partial f}{\partial \alpha_0} d\alpha_0
\end{equation}
which considering for simplicity $G \equiv 1$ give rise to:
\begin{equation}
    \frac{\partial \bar{m}}{\partial M} dM = \left ( \frac{1}{4\pi} f'\big |_{r=r_h}  \right ) \left ( \frac{d-2}{2}\right ) \left ( \frac{\alpha_1}{2} r_h^{d-3} + \alpha_2 r_h^{d-5} \right ) dr_h + \frac{d-2}{16\pi}r_h^{d-1} d\alpha_0
\end{equation}
Defining $\alpha_0=-2\Lambda$ with $\Lambda=-\dfrac{8\pi\Omega_{d-2}}{(d-2)(d-1)}p$, where $\Lambda$ and $p$ represent to the cosmological constant and the thermodynamics pressure in the extended phase space, the above equation can be rewritten as:
\begin{equation}
du=TdS+Vdp
\end{equation}
where $du$ corresponds to a local definition of the variation of energy at the horizon. From the last equation, it is direct to check that
\begin{align}
T&=\frac{1}{4\pi} f'\big |_{r=r_h} \\
S&= \frac{\alpha_1}{4} r_h^{d-2}+\frac{(d-2)\alpha_2}{2(d-4)}r_h^{d-4} \label{entropia} \\
V&= \frac{\Omega_{d-2}}{d-1}r_h^{d-1}
\end{align}
where the entropy has the form of the computed in reference \cite{Myers:1988ze} for the vacuum EGB case and the thermodynamic volume coincides with the geometric volume. So, we can note that under our redefinition of the first law of thermodynamics for a regular black hole in Einstein Gauss-Bonnet gravity are obtained the correct values of entropy and thermodynamics volume.

\subsubsection{\bf Conserved charge}

The mass of these solutions can be obtained using most of the methods described in the literature. See for instance \cite{Mora:2004rx}. Using \eqref{CondicionM}, the result is given by
\begin{equation}
    E = \left\{ \begin{array}{ll}
         M & \text{for  } d \text{ even}   \\
         M + E_v & \text{for } d \text{ odd}
    \end{array}\right.,
\end{equation}
where $E_v$ is the energy of the vacuum \cite{Mora:2004rx,Miskovic:2007mg}. 

\section{A toy model}
So far, it has not been proposed a specific form for the energy density and thus for the mass function \eqref{funciondemasa1}. This is necessary in order to test the horizon structure and the thermodynamic characteristics of the solution. It is worth noting that both the energy density, as the mass function, and the coupling constants $\alpha_k$ must satisfy the conditions exposed in the subsection \ref{ListaCriterios}. As a test of proof in this work, we use the energy density of reference \cite{Aros:2019quj}:
\begin{equation} \label{DensidadDeEnergia}
\rho(r)=\frac{(d-1)(d-2)}{8\pi} \frac{L^dM^2}{(L^dM+r^{d-1})^2}
\end{equation}
where $M$ is a constant related with the mass of units $[M]=\ell^{-1}$ and $L$ is a constant of units $[L]=\ell$. It is worth mentioning that this model of energy density is free of singularities and vanishes at infinity. 

This is called as an energy density with localized sources of matter because this takes local values for different values of the radial coordinate $r$.
Replacing \eqref{DensidadDeEnergia} in \eqref{funciondemasa1}, using $\dfrac{d-2}{8\pi}M$ as integration constant, and, after this, replacing in \eqref{funciondemasa2}:
\begin{equation}
m(r)=\frac{Mr^{d-1}}{L^dM+r^{d-1}}
\end{equation}
which satisfies the conditions \eqref{CondicionM} and \eqref{funcionmasa0}. Furthermore the units are $[m(r)]=\ell^{-1}$ as was mentioned in section \ref{SolucionGenerica}.

So, the solution is given by the equation \eqref{Solucion}, where, furthermore the coupling constants $\alpha_k$ must satisfy the conditions exposed in the subsection \ref{ListaCriterios}.

It is worth mentioning that, this energy density (and mass function) is only a test of proof. However, other models that satisfy the constraints of subsection \ref{ListaCriterios} could be used in future works for constructing RBH solutions in EGB gravity.

\subsection{Horizon structure} \label{HorizontesEstructura}
\begin{figure}[!h]
    \centering
    \begin{minipage}{0.5\linewidth}
        \centering
        \includegraphics[width=1.0\textwidth]{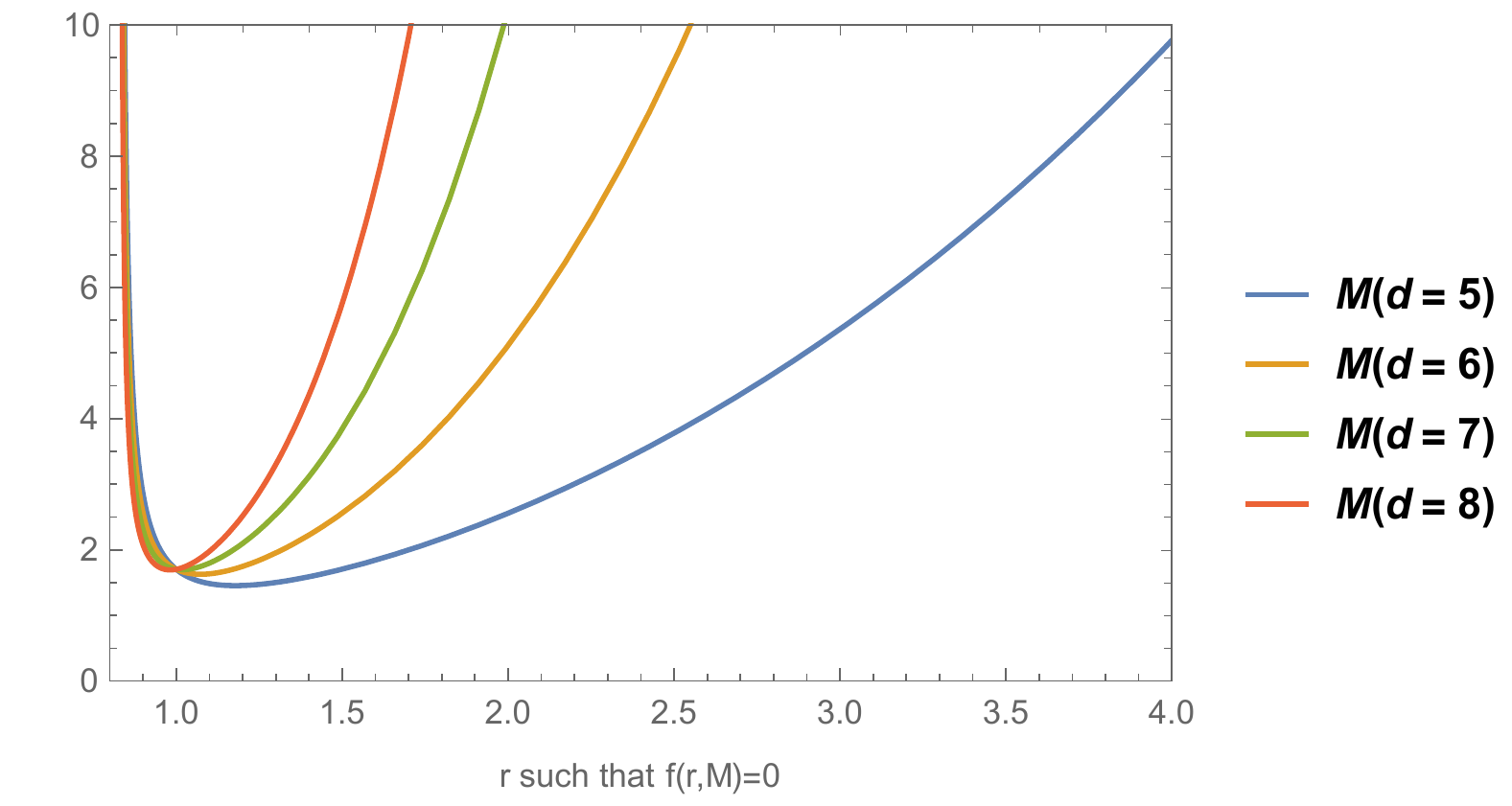}
        \label{Mplot}
    \end{minipage}\hfill
    \begin{minipage}{0.5\linewidth}
        \centering
        \includegraphics[width=1.1\textwidth]{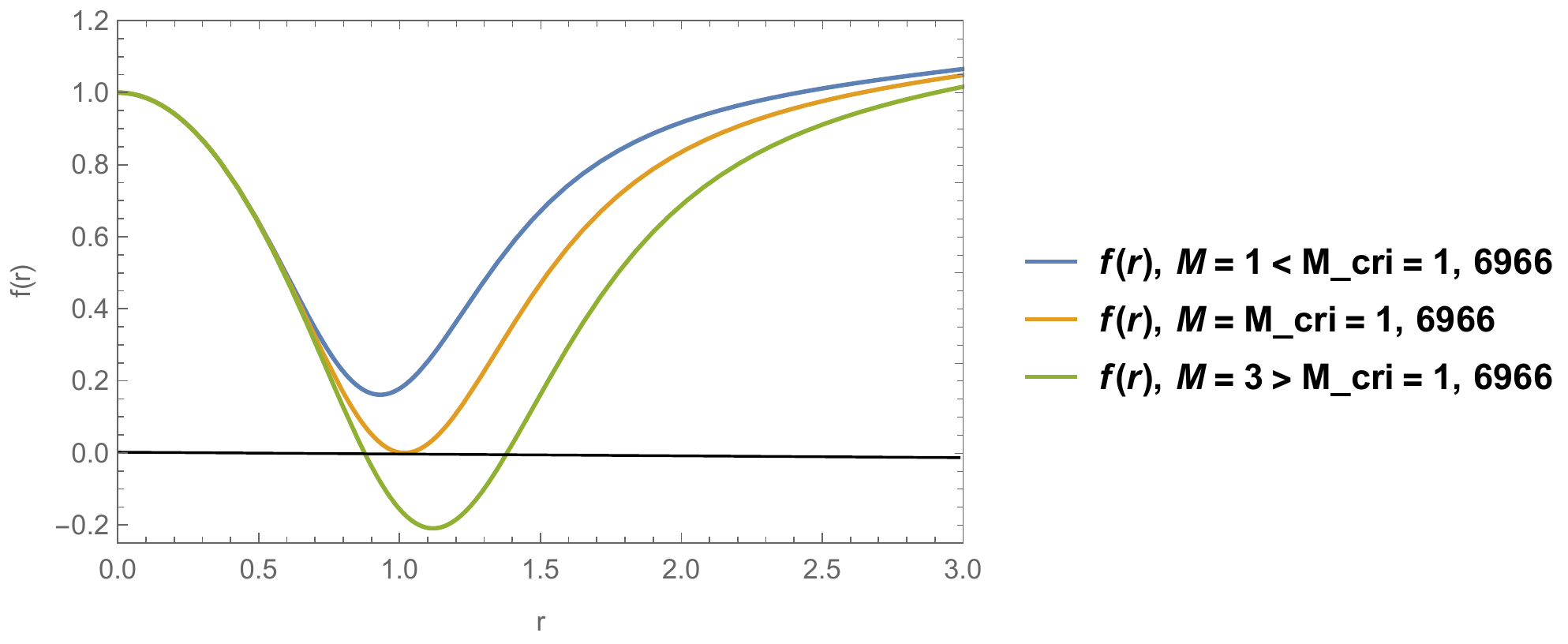}
              \label{fPlot}
    \end{minipage}
      \caption{Left figure: Mass parameter for $d=5,6,7,8$. Right figure: f(r) for $d=7$. In both figures $\alpha_2=1/4$, $\alpha_1=L=1$, $\alpha_0=0.01$ }
    \label{PlotHorizontes}
\end{figure}

We can see in the left side of figure \ref{PlotHorizontes} the behavior of the mass parameter $M$ such that $f(r,M)=0$ in the space of parameters for $d=5,6,7,8$ dimensions. For each curve, the values of the horizontal axis in the decreasing part of the left represent the internal horizon, $r_{in}$. Furthermore, the values of the horizontal axis in the increasing part of the right represent the black hole horizon, $r_h$. The minimum value of each curve corresponds to the critical value $M_{cri}$, where the internal and black hole horizons coincide, $r_{min}=r_{in}=r_h$, which represents the extremal black hole. We can notice that for larger dimensions the value of the radius of the extremal black hole, $r_{min}$, decreases, thus, the size of this latter becomes smaller for larger dimensions. The values of $r_{min}$ in the graphics are of the order of the unity, so, if the values of $r$ on the horizontal axis would correspond to Planck units, the extremal case would occur on the order of the Planck length, $r_{min} \approx \ell_p$, which has physical consequences as we will discuss below.

We can see the behavior of the function \eqref{Solucion} for $d=7$ in the right side of figure \ref{PlotHorizontes}. It is direct to check that this behavior is generic for other values of $d$. We can note that for values of $M<M_{cri}$ there are not horizons. For $M=M_{cri}$ the internal and black hole horizons coincide, $r_{min}=r_{in}=r_h$, which corresponds to the extremal black hole. For $M>M_{cri}$ there is the presence of the internal and black hole horizon, being the latter the larger one. It is worth mentioning that the number of horizons differs from the EGB vacuum case \cite{Boulware:1985wk}, which has only one. On the other hand, the structure of the horizons matches with RBH models in EGB with nonlinear electrodynamics sources \cite{Ovgun:2021ttv,Singh:2019wpu}.

 \subsection{Temperature}
\begin{figure}[!h]
    \centering
    \begin{minipage}{0.5\linewidth}
        \centering
        \includegraphics[width=1.0\textwidth]{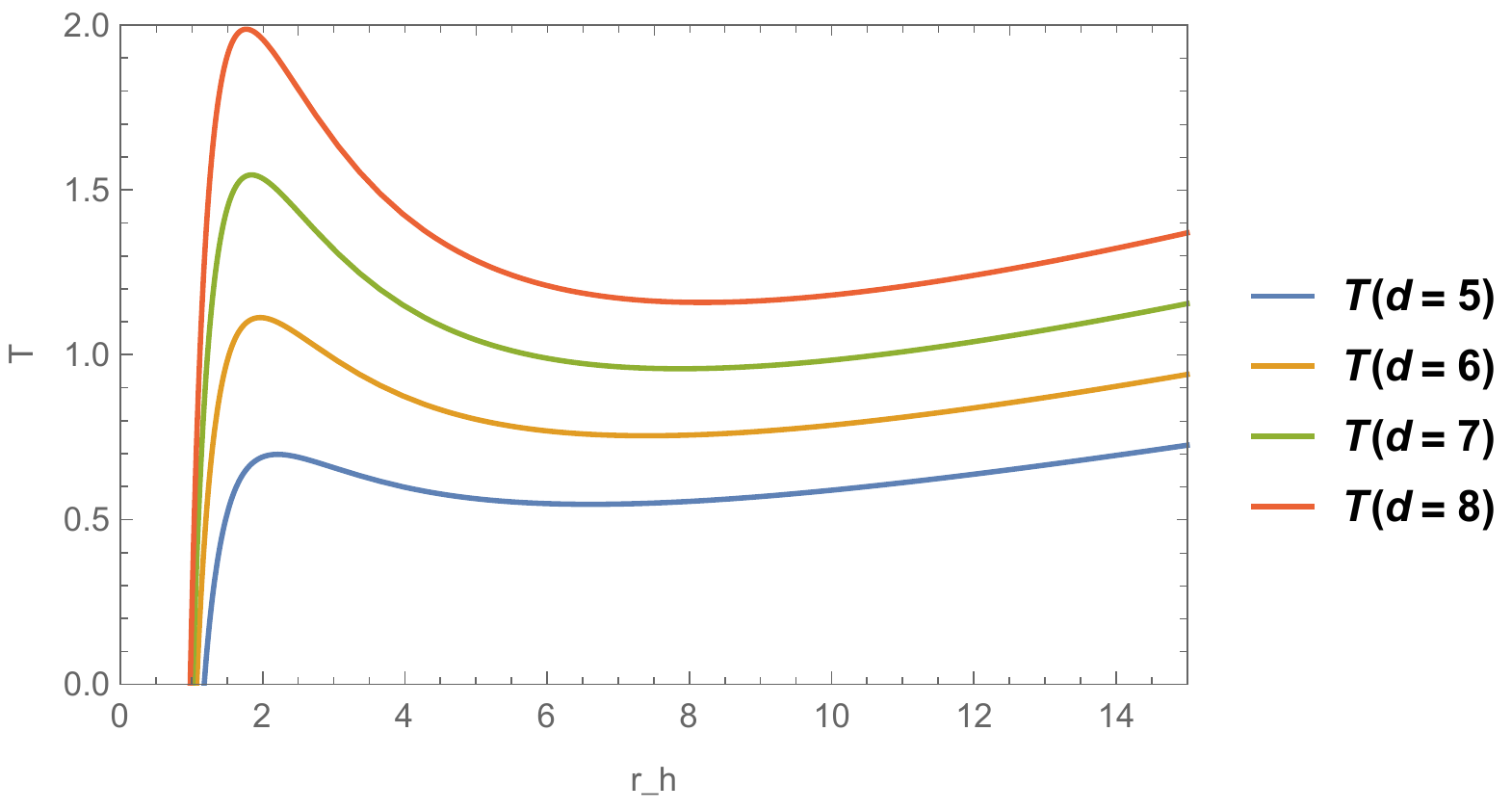}
        \label{Mplot}
    \end{minipage}\hfill
     \caption{Temperature for $d=5,6,7,8$ with $\alpha_2=1/4$, $\alpha_1=L=1$, $\alpha_0=0.01$ }
    \label{PlotTemperatura}
\end{figure}

We can see the behavior of the temperature for $d=5,6,7,8$ dimensions in the figure \ref{PlotTemperatura}. We can note that the point, $r_{min}$, where the mass parameter reaches a minimum ({\it i.e.} the internal and black hole horizons coincide, the left side of figure \ref{PlotTemperatura}), coincide with the point where the temperature vanishes. So, the temperature vanishes at the extremal black hole. It is well known the fact that the zero temperature point is associated with a black remnant, referred to as what is left behind once evaporation stops \cite{Adler:2001vs}. 

As it was mentioned in subsection \ref{HorizontesEstructura}, for larger dimensions the value of the radius of the extremal black hole, $r_{min}$, decreases, thus, the size of this latter becomes smaller for larger dimensions. Furthermore, also it was mentioned in subsection \ref{HorizontesEstructura} that, it is possible that the extremal black hole be reached at Planck scales. Then, the evaporation would stop once the horizon radius contracts up to a value close to the Planck length. This latter could be related to the apparition of quantum effects at this scale. 

We can also notice that for all the values of $d$, the behavior is generic. From left to right, the temperature increases from $T(r_{min})=0$ up to reaches a local maximum at $r_h=r_{cri1}$, after this decreases up to reach a minimum local at $r_h=r_{cri2}$. After this, from $r_h=r_{cri2}$ the temperature increases. It is worth mentioning that this behavior differs from the EGB vacuum one in \cite{Myers:1988ze}, whose derivative is negative. This behavior also differs from those of RBHs in EGB with nonlinear electrodynamics sources \cite{Ovgun:2021ttv,Singh:2019wpu}, whose derivatives have only one change of sign.

 \subsection{Heat capacity}

Following the standard thermodynamic definitions, this work calls second-order transition the discontinuity of the derivative of the Heat Capacity. In the same fashion, a stable phase corresponds to a range in the parameters where the Heat Capacity is positive, as this allows the system to reach thermal equilibrium with an external source. Unlike their EGB vacuum solution counterparts, whose Heat Capacities are always negative (indicating an unstable phase), for $d\ge 5$, as explained below, the presence of a source in this case generates regions, in the space of parameters, where the system can reach thermal equilibrium. The presence of these regions of stability and instability, although only formally, can be used to address, at least in general terms, the thermodynamic evolution of the system. Because of that, this can be used to foresee the evolution of the system from any configuration in the space of parameters.
 
The heat capacity can be computed as
\begin{equation} \label{CalorEspecifico}
    C = T \frac{dS}{dT}= T \left ( \frac{\partial S}{\partial r_h} \right) \left ( \frac{\partial T}{\partial r} \right)^{-1} \Big |_{r=r_h}
\end{equation}

From equation \eqref{entropia} it is direct to check that the sign of $\frac{\partial S}{\partial r_h}$ is always positive. So, given that the temperature is always positive, the value of the specific heat depends only on the value of the derivative $\frac{\partial T}{\partial r_h}$. From left to right in figure \ref{PlotTemperatura} we observe: a small--stable black hole, at the interval $[r_{min},r_{cri1}[$, where $\frac{\partial T}{\partial r_h}>0$. A unstable black hole, at the interval $]r_{cri1},r_{cri2}[$, where $\frac{\partial T}{\partial r_h}<0$. A large--stable black hole, after $r_h=r_{cri1}$, where $\frac{\partial T}{\partial r_h}>0$. It is worth mentioning that both in $r_h=r_{cri1}$ and in $r_h=r_{cri2}$ the derivative $\frac{\partial T}{\partial r_h}=0$ and the heat capacity diverges. So: in $r_h=r_{cri1}$ there is a second order phase transition between small stable--unstable black hole. In $r_h=r_{cri2}$ there is a second order phase transition between unstable--large stable black hole. It is worth stressing that the presence of matter fields in the energy--momentum tensor induces two phase transitions and two regions of stability. This latter differs from the vacuum EGB solution \cite{Myers:1988ze} where the specific heat is always negative without phase transition as occurs in Schwarzschild black hole. This behavior also differs from those of RBHs in EGB with nonlinear electrodynamics sources \cite{Ovgun:2021ttv,Singh:2019wpu}, where there is only one phase transition and only one region of stability.

 \section{Discussion and conclusion}
In this work, we have shown a new RBH solution in EGB gravity with localized sources of matter in the energy--momentum tensor. Due that the structure of the grounds state and the equations of motion differing from the GR and from particular cases of Lovelock gravity (as for example PL and LnF ), the construction of RBH with localized sources of matter in EGB gravity is not trivial. So, in section \ref{ListaCriterios} we have enunciated a list of constraints for the energy density and the coupling constants in order that the solution to be regular and have a correct limit $\alpha_2 \to 0$. It is worth mentioning that, although this work used a specific form for energy density, as a test of proof, these constraints could serve as a recipe for constructing several new RBH solutions in EGB gravity with several models of energy density. Particularly, in this work was chosen the energy density of reference \cite{Aros:2019quj}, given by the equation (34). Another energy density model that satisfies the same constraints of the subsection \ref{ListaCriterios}, corresponds to the higher dimensional generalization of the Dymnikova model \cite{Dymnikova:1992ux} discussed in reference \cite{Estrada:2019qsu}. Also could be tested the $d$-dimensional model of reference \cite{Hendi:2022opt}, or some $d$-dimensional generalization of the model proposed in reference \cite{Maluf:2022jjc}.

On the other hand, the usual version of the first law of thermodynamics, namely $dM=TdS+PdV$ is not valid for RBH due to the presence of matter fields in the energy--momentum tensor \cite{Ma:2014qma}. This is because the usual form of the first law leads to incorrect values of entropy and volume. So, for the EGB gravity in $d$ dimensions, we have proposed a form for the first law of thermodynamics for RBH, using a local definition of the variation of energy at the horizon by imposing constraints on the evolution along the space parameters. Remarkably, our form for the first law leads to the correct value of entropy, which coincides with the vacuum case \cite{Myers:1988ze} and the thermodynamics volume coincides with the geometric volume. 

By using the energy density in reference \cite{Aros:2019quj}, we have studied the horizons structure and the thermodynamics behavior of the solutions. There is a critical value of the mass parameter $M_{cri}$ such that: for values of $M<M_{cri}$ there are no horizons. For $M=M_{cri}$ the internal and black hole horizons coincide, $r_{min}=r_{in}=r_h$, which corresponds to the extremal black hole. For $M>M_{cri}$ there is the presence of the internal and black hole horizon, being the latter the larger one. 

We can notice that for larger dimensions the value of the radius of the extremal black hole, $r_{min}$, decreases, thus, the size of this latter becomes smaller for larger dimensions. The values of $r_{min}$ in the graphics can be of the order of the unity, so, if the values of $r$ on the horizontal axis would correspond to Planck units, the extremal case would occur on the order of the Planck length, $r_{min} \approx \ell_p$. In the extremal black hole, the temperature vanishes, which is associated with a black remnant,
referred to as what is left behind once evaporation stops. Thus it is possible that, if the radius of the extremal black hole is of order of the Planck length,  thus the evaporation would stop once the horizon radius contracts up to a value close to the Planck length. This latter could be related to the apparition of quantum effects at this scale. The latter can be justified as an effect of the Generalized Uncertainty Principle (GUP), which applied at Planck scales should prevent total evaporation in the fashion as the uncertainty principle prevents the collapse of the Hydrogen atom at atomic scales. See \cite{Adler:2001vs} for general discussion and \cite{Kuntz:2019gka} for a discussion about the emission of fermions. Once one had accepted the idea of a $T=0$ remnant as the final stage of evolution one can speculate on the physical consequence of this. First, one speculates that a hypothetical group of these $T=0$ remnants is a suitable candidate for cold dark matter as they have very small scattering effective sections, thus they behave essentially as non-interacting particles. However, even in this scenario of a $T=0$ remnant, one must mention \cite{Ward:2006hf} where it is proposed that the remnant is expected to decay into $n$-body final states. In this case, this leads to Planck-scale cosmic rays. See also the references \cite{Zhang:2020qxw,Ali:2015tva}.

It is worth stressing that the presence of matter fields in the energy--momentum tensor induces two phase transitions, between small stable--unstable and between unstable--large stable black holes. So, there are two regions of stability. This differs from the vacuum EGB solution \cite{Myers:1988ze} where the specific heat is always negative without phase transition as occurs in Schwarzschild black hole.

\section*{Acknowledgements}
This work of RA was partially funded through FONDECYT-Chile 1220335. Milko Estrada is funded by ANID , FONDECYT de Iniciaci\'on en Investigación 2023, Folio 11230247.

\bibliography{mybib} 

%merlin.mbs apsrev4-1.bst 2010-07-25 4.21a (PWD, AO, DPC) hacked
%Control: key (0)
%Control: author (0) dotless jnrlst
%Control: editor formatted (1) identically to author
%Control: production of article title (0) allowed
%Control: page (1) range
%Control: year (0) verbatim
%Control: production of eprint (0) enabled
\begin{thebibliography}{41}%
\makeatletter
\providecommand \@ifxundefined [1]{%
 \@ifx{#1\undefined}
}%
\providecommand \@ifnum [1]{%
 \ifnum #1\expandafter \@firstoftwo
 \else \expandafter \@secondoftwo
 \fi
}%
\providecommand \@ifx [1]{%
 \ifx #1\expandafter \@firstoftwo
 \else \expandafter \@secondoftwo
 \fi
}%
\providecommand \natexlab [1]{#1}%
\providecommand \enquote  [1]{``#1''}%
\providecommand \bibnamefont  [1]{#1}%
\providecommand \bibfnamefont [1]{#1}%
\providecommand \citenamefont [1]{#1}%
\providecommand \href@noop [0]{\@secondoftwo}%
\providecommand \href [0]{\begingroup \@sanitize@url \@href}%
\providecommand \@href[1]{\@@startlink{#1}\@@href}%
\providecommand \@@href[1]{\endgroup#1\@@endlink}%
\providecommand \@sanitize@url [0]{\catcode `\\12\catcode `\$12\catcode
  `\&12\catcode `\#12\catcode `\^12\catcode `\_12\catcode `\%12\relax}%
\providecommand \@@startlink[1]{}%
\providecommand \@@endlink[0]{}%
\providecommand \url  [0]{\begingroup\@sanitize@url \@url }%
\providecommand \@url [1]{\endgroup\@href {#1}{\urlprefix }}%
\providecommand \urlprefix  [0]{URL }%
\providecommand \Eprint [0]{\href }%
\providecommand \doibase [0]{http://dx.doi.org/}%
\providecommand \selectlanguage [0]{\@gobble}%
\providecommand \bibinfo  [0]{\@secondoftwo}%
\providecommand \bibfield  [0]{\@secondoftwo}%
\providecommand \translation [1]{[#1]}%
\providecommand \BibitemOpen [0]{}%
\providecommand \bibitemStop [0]{}%
\providecommand \bibitemNoStop [0]{.\EOS\space}%
\providecommand \EOS [0]{\spacefactor3000\relax}%
\providecommand \BibitemShut  [1]{\csname bibitem#1\endcsname}%
\let\auto@bib@innerbib\@empty
%</preamble>
\bibitem [{\citenamefont {Lovelock}(1971)}]{Lovelock:1971yv}%
  \BibitemOpen
  \bibfield  {author} {\bibinfo {author} {\bibfnamefont {D.}~\bibnamefont
  {Lovelock}},\ }\bibfield  {title} {\enquote {\bibinfo {title} {{The Einstein
  tensor and its generalizations}},}\ }\href {\doibase 10.1063/1.1665613}
  {\bibfield  {journal} {\bibinfo  {journal} {J. Math. Phys.}\ }\textbf
  {\bibinfo {volume} {12}},\ \bibinfo {pages} {498--501} (\bibinfo {year}
  {1971})}\BibitemShut {NoStop}%
\bibitem [{\citenamefont {Oikonomou}(2021)}]{Oikonomou:2021kql}%
  \BibitemOpen
  \bibfield  {author} {\bibinfo {author} {\bibfnamefont {V.~K.}\ \bibnamefont
  {Oikonomou}},\ }\bibfield  {title} {\enquote {\bibinfo {title} {{A refined
  Einstein\textendash{}Gauss\textendash{}Bonnet inflationary theoretical
  framework}},}\ }\href {\doibase 10.1088/1361-6382/ac2168} {\bibfield
  {journal} {\bibinfo  {journal} {Class. Quant. Grav.}\ }\textbf {\bibinfo
  {volume} {38}},\ \bibinfo {pages} {195025} (\bibinfo {year} {2021})},\
  \Eprint {http://arxiv.org/abs/2108.10460} {arXiv:2108.10460 [gr-qc]}
  \BibitemShut {NoStop}%
\bibitem [{\citenamefont {Odintsov}\ and\ \citenamefont
  {Oikonomou}(2020)}]{Odintsov:2020zkl}%
  \BibitemOpen
  \bibfield  {author} {\bibinfo {author} {\bibfnamefont {S.~D.}\ \bibnamefont
  {Odintsov}}\ and\ \bibinfo {author} {\bibfnamefont {V.~K.}\ \bibnamefont
  {Oikonomou}},\ }\bibfield  {title} {\enquote {\bibinfo {title} {{Swampland
  implications of GW170817-compatible Einstein-Gauss-Bonnet gravity}},}\ }\href
  {\doibase 10.1016/j.physletb.2020.135437} {\bibfield  {journal} {\bibinfo
  {journal} {Phys. Lett. B}\ }\textbf {\bibinfo {volume} {805}},\ \bibinfo
  {pages} {135437} (\bibinfo {year} {2020})},\ \Eprint
  {http://arxiv.org/abs/2004.00479} {arXiv:2004.00479 [gr-qc]} \BibitemShut
  {NoStop}%
\bibitem [{\citenamefont {Glavan}\ and\ \citenamefont
  {Lin}(2020)}]{Glavan:2019inb}%
  \BibitemOpen
  \bibfield  {author} {\bibinfo {author} {\bibfnamefont {Dra\v{z}en}\
  \bibnamefont {Glavan}}\ and\ \bibinfo {author} {\bibfnamefont {Chunshan}\
  \bibnamefont {Lin}},\ }\bibfield  {title} {\enquote {\bibinfo {title}
  {{Einstein-Gauss-Bonnet Gravity in Four-Dimensional Spacetime}},}\ }\href
  {\doibase 10.1103/PhysRevLett.124.081301} {\bibfield  {journal} {\bibinfo
  {journal} {Phys. Rev. Lett.}\ }\textbf {\bibinfo {volume} {124}},\ \bibinfo
  {pages} {081301} (\bibinfo {year} {2020})},\ \Eprint
  {http://arxiv.org/abs/1905.03601} {arXiv:1905.03601 [gr-qc]} \BibitemShut
  {NoStop}%
\bibitem [{\citenamefont {Lu}\ and\ \citenamefont {Pang}(2020)}]{Lu:2020iav}%
  \BibitemOpen
  \bibfield  {author} {\bibinfo {author} {\bibfnamefont {H.}~\bibnamefont
  {Lu}}\ and\ \bibinfo {author} {\bibfnamefont {Yi}~\bibnamefont {Pang}},\
  }\bibfield  {title} {\enquote {\bibinfo {title} {{Horndeski gravity as $D
  \rightarrow 4$ limit of Gauss-Bonnet}},}\ }\href {\doibase
  10.1016/j.physletb.2020.135717} {\bibfield  {journal} {\bibinfo  {journal}
  {Phys. Lett. B}\ }\textbf {\bibinfo {volume} {809}},\ \bibinfo {pages}
  {135717} (\bibinfo {year} {2020})},\ \Eprint
  {http://arxiv.org/abs/2003.11552} {arXiv:2003.11552 [gr-qc]} \BibitemShut
  {NoStop}%
\bibitem [{\citenamefont {Fernandes}\ \emph {et~al.}(2020)\citenamefont
  {Fernandes}, \citenamefont {Carrilho}, \citenamefont {Clifton},\ and\
  \citenamefont {Mulryne}}]{Fernandes:2020nbq}%
  \BibitemOpen
  \bibfield  {author} {\bibinfo {author} {\bibfnamefont {Pedro G.~S.}\
  \bibnamefont {Fernandes}}, \bibinfo {author} {\bibfnamefont {Pedro}\
  \bibnamefont {Carrilho}}, \bibinfo {author} {\bibfnamefont {Timothy}\
  \bibnamefont {Clifton}}, \ and\ \bibinfo {author} {\bibfnamefont {David~J.}\
  \bibnamefont {Mulryne}},\ }\bibfield  {title} {\enquote {\bibinfo {title}
  {{Derivation of Regularized Field Equations for the Einstein-Gauss-Bonnet
  Theory in Four Dimensions}},}\ }\href {\doibase 10.1103/PhysRevD.102.024025}
  {\bibfield  {journal} {\bibinfo  {journal} {Phys. Rev. D}\ }\textbf {\bibinfo
  {volume} {102}},\ \bibinfo {pages} {024025} (\bibinfo {year} {2020})},\
  \Eprint {http://arxiv.org/abs/2004.08362} {arXiv:2004.08362 [gr-qc]}
  \BibitemShut {NoStop}%
\bibitem [{\citenamefont {Fernandes}(2021)}]{Fernandes:2021dsb}%
  \BibitemOpen
  \bibfield  {author} {\bibinfo {author} {\bibfnamefont {Pedro G.~S.}\
  \bibnamefont {Fernandes}},\ }\bibfield  {title} {\enquote {\bibinfo {title}
  {{Gravity with a generalized conformal scalar field: theory and
  solutions}},}\ }\href {\doibase 10.1103/PhysRevD.103.104065} {\bibfield
  {journal} {\bibinfo  {journal} {Phys. Rev. D}\ }\textbf {\bibinfo {volume}
  {103}},\ \bibinfo {pages} {104065} (\bibinfo {year} {2021})},\ \Eprint
  {http://arxiv.org/abs/2105.04687} {arXiv:2105.04687 [gr-qc]} \BibitemShut
  {NoStop}%
\bibitem [{\citenamefont {\"Ovg\"un}(2021)}]{Ovgun:2021ttv}%
  \BibitemOpen
  \bibfield  {author} {\bibinfo {author} {\bibfnamefont {A.}~\bibnamefont
  {\"Ovg\"un}},\ }\bibfield  {title} {\enquote {\bibinfo {title} {{Black hole
  with confining electric potential in scalar-tensor description of regularized
  4-dimensional Einstein-Gauss-Bonnet gravity}},}\ }\href {\doibase
  10.1016/j.physletb.2021.136517} {\bibfield  {journal} {\bibinfo  {journal}
  {Phys. Lett. B}\ }\textbf {\bibinfo {volume} {820}},\ \bibinfo {pages}
  {136517} (\bibinfo {year} {2021})},\ \Eprint
  {http://arxiv.org/abs/2105.05035} {arXiv:2105.05035 [gr-qc]} \BibitemShut
  {NoStop}%
\bibitem [{\citenamefont {Aoki}\ \emph {et~al.}(2020)\citenamefont {Aoki},
  \citenamefont {Gorji},\ and\ \citenamefont {Mukohyama}}]{Aoki:2020lig}%
  \BibitemOpen
  \bibfield  {author} {\bibinfo {author} {\bibfnamefont {Katsuki}\ \bibnamefont
  {Aoki}}, \bibinfo {author} {\bibfnamefont {Mohammad~Ali}\ \bibnamefont
  {Gorji}}, \ and\ \bibinfo {author} {\bibfnamefont {Shinji}\ \bibnamefont
  {Mukohyama}},\ }\bibfield  {title} {\enquote {\bibinfo {title} {{A consistent
  theory of $D \to 4$ Einstein-Gauss-Bonnet gravity}},}\ }\href {\doibase
  10.1016/j.physletb.2020.135843} {\bibfield  {journal} {\bibinfo  {journal}
  {Phys. Lett. B}\ }\textbf {\bibinfo {volume} {810}},\ \bibinfo {pages}
  {135843} (\bibinfo {year} {2020})},\ \Eprint
  {http://arxiv.org/abs/2005.03859} {arXiv:2005.03859 [gr-qc]} \BibitemShut
  {NoStop}%
\bibitem [{\citenamefont {Abbott}\ \emph {et~al.}(2016)\citenamefont {Abbott}
  \emph {et~al.}}]{LIGOScientific:2016aoc}%
  \BibitemOpen
  \bibfield  {author} {\bibinfo {author} {\bibfnamefont {B.~P.}\ \bibnamefont
  {Abbott}} \emph {et~al.} (\bibinfo {collaboration} {LIGO Scientific,
  Virgo}),\ }\bibfield  {title} {\enquote {\bibinfo {title} {{Observation of
  Gravitational Waves from a Binary Black Hole Merger}},}\ }\href {\doibase
  10.1103/PhysRevLett.116.061102} {\bibfield  {journal} {\bibinfo  {journal}
  {Phys. Rev. Lett.}\ }\textbf {\bibinfo {volume} {116}},\ \bibinfo {pages}
  {061102} (\bibinfo {year} {2016})},\ \Eprint
  {http://arxiv.org/abs/1602.03837} {arXiv:1602.03837 [gr-qc]} \BibitemShut
  {NoStop}%
\bibitem [{\citenamefont {Bardeen}(1968)}]{bardeen}%
  \BibitemOpen
  \bibfield  {author} {\bibinfo {author} {\bibfnamefont {James}\ \bibnamefont
  {Bardeen}},\ }\href@noop {} {\bibfield  {journal} {\bibinfo  {journal}
  {Proceedings of the International Conference GR5, Tbilisi USSR}\ } (\bibinfo
  {year} {1968})}\BibitemShut {NoStop}%
\bibitem [{\citenamefont {Ayon-Beato}\ and\ \citenamefont
  {Garcia}(2000)}]{Ayon-Beato:2000mjt}%
  \BibitemOpen
  \bibfield  {author} {\bibinfo {author} {\bibfnamefont {Eloy}\ \bibnamefont
  {Ayon-Beato}}\ and\ \bibinfo {author} {\bibfnamefont {Alberto}\ \bibnamefont
  {Garcia}},\ }\bibfield  {title} {\enquote {\bibinfo {title} {{The Bardeen
  model as a nonlinear magnetic monopole}},}\ }\href {\doibase
  10.1016/S0370-2693(00)01125-4} {\bibfield  {journal} {\bibinfo  {journal}
  {Phys. Lett. B}\ }\textbf {\bibinfo {volume} {493}},\ \bibinfo {pages}
  {149--152} (\bibinfo {year} {2000})},\ \Eprint
  {http://arxiv.org/abs/gr-qc/0009077} {arXiv:gr-qc/0009077} \BibitemShut
  {NoStop}%
\bibitem [{\citenamefont {Singh}\ \emph {et~al.}(2020)\citenamefont {Singh},
  \citenamefont {Ghosh},\ and\ \citenamefont {Maharaj}}]{Singh:2019wpu}%
  \BibitemOpen
  \bibfield  {author} {\bibinfo {author} {\bibfnamefont {Dharm~Veer}\
  \bibnamefont {Singh}}, \bibinfo {author} {\bibfnamefont {Sushant~G.}\
  \bibnamefont {Ghosh}}, \ and\ \bibinfo {author} {\bibfnamefont {Sunil~D.}\
  \bibnamefont {Maharaj}},\ }\bibfield  {title} {\enquote {\bibinfo {title}
  {{Bardeen-like regular black holes in $5D$ Einstein-Gauss-Bonnet gravity}},}\
  }\href {\doibase 10.1016/j.aop.2019.168025} {\bibfield  {journal} {\bibinfo
  {journal} {Annals Phys.}\ }\textbf {\bibinfo {volume} {412}},\ \bibinfo
  {pages} {168025} (\bibinfo {year} {2020})},\ \Eprint
  {http://arxiv.org/abs/1911.11054} {arXiv:1911.11054 [gr-qc]} \BibitemShut
  {NoStop}%
\bibitem [{\citenamefont {Sorokin}(2022)}]{Sorokin:2021tge}%
  \BibitemOpen
  \bibfield  {author} {\bibinfo {author} {\bibfnamefont {Dmitri~P.}\
  \bibnamefont {Sorokin}},\ }\bibfield  {title} {\enquote {\bibinfo {title}
  {{Introductory Notes on Non-linear Electrodynamics and its Applications}},}\
  }\href {\doibase 10.1002/prop.202200092} {\bibfield  {journal} {\bibinfo
  {journal} {Fortsch. Phys.}\ }\textbf {\bibinfo {volume} {70}},\ \bibinfo
  {pages} {2200092} (\bibinfo {year} {2022})},\ \Eprint
  {http://arxiv.org/abs/2112.12118} {arXiv:2112.12118 [hep-th]} \BibitemShut
  {NoStop}%
\bibitem [{\citenamefont {Mkrtchyan}\ and\ \citenamefont
  {Svazas}(2022)}]{Mkrtchyan:2022ulc}%
  \BibitemOpen
  \bibfield  {author} {\bibinfo {author} {\bibfnamefont {Karapet}\ \bibnamefont
  {Mkrtchyan}}\ and\ \bibinfo {author} {\bibfnamefont {Mantas}\ \bibnamefont
  {Svazas}},\ }\bibfield  {title} {\enquote {\bibinfo {title} {{Solutions in
  Nonlinear Electrodynamics and their double copy regular black holes}},}\
  }\href {\doibase 10.1007/JHEP09(2022)012} {\bibfield  {journal} {\bibinfo
  {journal} {JHEP}\ }\textbf {\bibinfo {volume} {09}},\ \bibinfo {pages} {012}
  (\bibinfo {year} {2022})},\ \Eprint {http://arxiv.org/abs/2205.14187}
  {arXiv:2205.14187 [hep-th]} \BibitemShut {NoStop}%
\bibitem [{\citenamefont {Malafarina}\ and\ \citenamefont
  {Toshmatov}(2022)}]{Malafarina:2022oka}%
  \BibitemOpen
  \bibfield  {author} {\bibinfo {author} {\bibfnamefont {Daniele}\ \bibnamefont
  {Malafarina}}\ and\ \bibinfo {author} {\bibfnamefont {Bobir}\ \bibnamefont
  {Toshmatov}},\ }\bibfield  {title} {\enquote {\bibinfo {title} {{Connection
  between regular black holes in nonlinear electrodynamics and semiclassical
  dust collapse}},}\ }\href {\doibase 10.1103/PhysRevD.105.L121502} {\bibfield
  {journal} {\bibinfo  {journal} {Phys. Rev. D}\ }\textbf {\bibinfo {volume}
  {105}},\ \bibinfo {pages} {L121502} (\bibinfo {year} {2022})},\ \Eprint
  {http://arxiv.org/abs/2204.04025} {arXiv:2204.04025 [gr-qc]} \BibitemShut
  {NoStop}%
\bibitem [{\citenamefont {Sajadi}\ and\ \citenamefont
  {Riazi}(2017)}]{Sajadi:2017glu}%
  \BibitemOpen
  \bibfield  {author} {\bibinfo {author} {\bibfnamefont {S.~N.}\ \bibnamefont
  {Sajadi}}\ and\ \bibinfo {author} {\bibfnamefont {N.}~\bibnamefont {Riazi}},\
  }\bibfield  {title} {\enquote {\bibinfo {title} {{Nonlinear electrodynamics
  and regular black holes}},}\ }\href {\doibase 10.1007/s10714-017-2209-8}
  {\bibfield  {journal} {\bibinfo  {journal} {Gen. Rel. Grav.}\ }\textbf
  {\bibinfo {volume} {49}},\ \bibinfo {pages} {45} (\bibinfo {year} {2017})},\
  \bibinfo {note} {[Erratum: Gen.Rel.Grav. 52, 18 (2020)]}\BibitemShut
  {NoStop}%
\bibitem [{\citenamefont {Sajadi}\ \emph {et~al.}(2019)\citenamefont {Sajadi},
  \citenamefont {Riazi},\ and\ \citenamefont {Hendi}}]{Sajadi:2019hzo}%
  \BibitemOpen
  \bibfield  {author} {\bibinfo {author} {\bibfnamefont {S.~N.}\ \bibnamefont
  {Sajadi}}, \bibinfo {author} {\bibfnamefont {N.}~\bibnamefont {Riazi}}, \
  and\ \bibinfo {author} {\bibfnamefont {S.~H.}\ \bibnamefont {Hendi}},\
  }\bibfield  {title} {\enquote {\bibinfo {title} {{Dynamical and thermal
  stabilities of nonlinearly charged AdS black holes}},}\ }\href {\doibase
  10.1140/epjc/s10052-019-7272-8} {\bibfield  {journal} {\bibinfo  {journal}
  {Eur. Phys. J. C}\ }\textbf {\bibinfo {volume} {79}},\ \bibinfo {pages} {775}
  (\bibinfo {year} {2019})},\ \Eprint {http://arxiv.org/abs/2003.13472}
  {arXiv:2003.13472 [gr-qc]} \BibitemShut {NoStop}%
\bibitem [{\citenamefont {Hendi}\ \emph {et~al.}(2021)\citenamefont {Hendi},
  \citenamefont {Sajadi},\ and\ \citenamefont {Khademi}}]{Hendi:2020knv}%
  \BibitemOpen
  \bibfield  {author} {\bibinfo {author} {\bibfnamefont {S.~H.}\ \bibnamefont
  {Hendi}}, \bibinfo {author} {\bibfnamefont {S.~N.}\ \bibnamefont {Sajadi}}, \
  and\ \bibinfo {author} {\bibfnamefont {M.}~\bibnamefont {Khademi}},\
  }\bibfield  {title} {\enquote {\bibinfo {title} {{Physical properties of a
  regular rotating black hole: Thermodynamics, stability, and quasinormal
  modes}},}\ }\href {\doibase 10.1103/PhysRevD.103.064016} {\bibfield
  {journal} {\bibinfo  {journal} {Phys. Rev. D}\ }\textbf {\bibinfo {volume}
  {103}},\ \bibinfo {pages} {064016} (\bibinfo {year} {2021})},\ \Eprint
  {http://arxiv.org/abs/2006.11575} {arXiv:2006.11575 [gr-qc]} \BibitemShut
  {NoStop}%
\bibitem [{\citenamefont {Sajadi}\ \emph {et~al.}(2021)\citenamefont {Sajadi},
  \citenamefont {Jahani~Poshteh},\ and\ \citenamefont
  {Hendi}}]{Sajadi:2021ilt}%
  \BibitemOpen
  \bibfield  {author} {\bibinfo {author} {\bibfnamefont {S.~N.}\ \bibnamefont
  {Sajadi}}, \bibinfo {author} {\bibfnamefont {M.~B.}\ \bibnamefont
  {Jahani~Poshteh}}, \ and\ \bibinfo {author} {\bibfnamefont {S.~H.}\
  \bibnamefont {Hendi}},\ }\bibfield  {title} {\enquote {\bibinfo {title}
  {{Instability of regular electric black hole}},}\ }\href {\doibase
  10.1016/j.nuclphysb.2021.115567} {\bibfield  {journal} {\bibinfo  {journal}
  {Nucl. Phys. B}\ }\textbf {\bibinfo {volume} {972}},\ \bibinfo {pages}
  {115567} (\bibinfo {year} {2021})}\BibitemShut {NoStop}%
\bibitem [{\citenamefont {Aros}\ and\ \citenamefont
  {Estrada}(2019)}]{Aros:2019quj}%
  \BibitemOpen
  \bibfield  {author} {\bibinfo {author} {\bibfnamefont {Rodrigo}\ \bibnamefont
  {Aros}}\ and\ \bibinfo {author} {\bibfnamefont {Milko}\ \bibnamefont
  {Estrada}},\ }\bibfield  {title} {\enquote {\bibinfo {title} {{Regular black
  holes and its thermodynamics in Lovelock gravity}},}\ }\href {\doibase
  10.1140/epjc/s10052-019-6783-7} {\bibfield  {journal} {\bibinfo  {journal}
  {Eur. Phys. J. C}\ }\textbf {\bibinfo {volume} {79}},\ \bibinfo {pages} {259}
  (\bibinfo {year} {2019})},\ \Eprint {http://arxiv.org/abs/1901.08724}
  {arXiv:1901.08724 [gr-qc]} \BibitemShut {NoStop}%
\bibitem [{\citenamefont {Cai}\ and\ \citenamefont {Ohta}(2006)}]{Cai:2006pq}%
  \BibitemOpen
  \bibfield  {author} {\bibinfo {author} {\bibfnamefont {Rong-Gen}\
  \bibnamefont {Cai}}\ and\ \bibinfo {author} {\bibfnamefont {Nobuyoshi}\
  \bibnamefont {Ohta}},\ }\bibfield  {title} {\enquote {\bibinfo {title}
  {{Black Holes in Pure Lovelock Gravities}},}\ }\href {\doibase
  10.1103/PhysRevD.74.064001} {\bibfield  {journal} {\bibinfo  {journal} {Phys.
  Rev. D}\ }\textbf {\bibinfo {volume} {74}},\ \bibinfo {pages} {064001}
  (\bibinfo {year} {2006})},\ \Eprint {http://arxiv.org/abs/hep-th/0604088}
  {arXiv:hep-th/0604088} \BibitemShut {NoStop}%
\bibitem [{\citenamefont {Crisostomo}\ \emph {et~al.}(2000)\citenamefont
  {Crisostomo}, \citenamefont {Troncoso},\ and\ \citenamefont
  {Zanelli}}]{Crisostomo:2000bb}%
  \BibitemOpen
  \bibfield  {author} {\bibinfo {author} {\bibfnamefont {Juan}\ \bibnamefont
  {Crisostomo}}, \bibinfo {author} {\bibfnamefont {Ricardo}\ \bibnamefont
  {Troncoso}}, \ and\ \bibinfo {author} {\bibfnamefont {Jorge}\ \bibnamefont
  {Zanelli}},\ }\bibfield  {title} {\enquote {\bibinfo {title} {{Black hole
  scan}},}\ }\href {\doibase 10.1103/PhysRevD.62.084013} {\bibfield  {journal}
  {\bibinfo  {journal} {Phys. Rev. D}\ }\textbf {\bibinfo {volume} {62}},\
  \bibinfo {pages} {084013} (\bibinfo {year} {2000})},\ \Eprint
  {http://arxiv.org/abs/hep-th/0003271} {arXiv:hep-th/0003271} \BibitemShut
  {NoStop}%
\bibitem [{\citenamefont {De~Lorenzo}\ \emph {et~al.}(2015)\citenamefont
  {De~Lorenzo}, \citenamefont {Pacilio}, \citenamefont {Rovelli},\ and\
  \citenamefont {Speziale}}]{DeLorenzo:2014pta}%
  \BibitemOpen
  \bibfield  {author} {\bibinfo {author} {\bibfnamefont {Tommaso}\ \bibnamefont
  {De~Lorenzo}}, \bibinfo {author} {\bibfnamefont {Costantino}\ \bibnamefont
  {Pacilio}}, \bibinfo {author} {\bibfnamefont {Carlo}\ \bibnamefont
  {Rovelli}}, \ and\ \bibinfo {author} {\bibfnamefont {Simone}\ \bibnamefont
  {Speziale}},\ }\bibfield  {title} {\enquote {\bibinfo {title} {{On the
  Effective Metric of a Planck Star}},}\ }\href {\doibase
  10.1007/s10714-015-1882-8} {\bibfield  {journal} {\bibinfo  {journal} {Gen.
  Rel. Grav.}\ }\textbf {\bibinfo {volume} {47}},\ \bibinfo {pages} {41}
  (\bibinfo {year} {2015})},\ \Eprint {http://arxiv.org/abs/1412.6015}
  {arXiv:1412.6015 [gr-qc]} \BibitemShut {NoStop}%
\bibitem [{\citenamefont {Estrada}\ and\ \citenamefont
  {Tello-Ortiz}(2021)}]{Estrada:2020tbz}%
  \BibitemOpen
  \bibfield  {author} {\bibinfo {author} {\bibfnamefont {Milko}\ \bibnamefont
  {Estrada}}\ and\ \bibinfo {author} {\bibfnamefont {Francisco}\ \bibnamefont
  {Tello-Ortiz}},\ }\bibfield  {title} {\enquote {\bibinfo {title} {{A new
  model of regular black hole in (2+1) dimensions}},}\ }\href {\doibase
  10.1209/0295-5075/ac0ed0} {\bibfield  {journal} {\bibinfo  {journal} {EPL}\
  }\textbf {\bibinfo {volume} {135}},\ \bibinfo {pages} {20001} (\bibinfo
  {year} {2021})},\ \Eprint {http://arxiv.org/abs/2012.05068} {arXiv:2012.05068
  [gr-qc]} \BibitemShut {NoStop}%
\bibitem [{\citenamefont {Maluf}\ \emph {et~al.}(2022)\citenamefont {Maluf},
  \citenamefont {Muniz}, \citenamefont {Santos},\ and\ \citenamefont
  {Estrada}}]{Maluf:2022jjc}%
  \BibitemOpen
  \bibfield  {author} {\bibinfo {author} {\bibfnamefont {R.~V.}\ \bibnamefont
  {Maluf}}, \bibinfo {author} {\bibfnamefont {C.~R.}\ \bibnamefont {Muniz}},
  \bibinfo {author} {\bibfnamefont {A.~C.~L.}\ \bibnamefont {Santos}}, \ and\
  \bibinfo {author} {\bibfnamefont {Milko}\ \bibnamefont {Estrada}},\
  }\bibfield  {title} {\enquote {\bibinfo {title} {{A new class of regular
  black hole solutions with quasi-localized sources of matter in (2+1)
  dimensions}},}\ }\href {\doibase 10.1016/j.physletb.2022.137581} {\bibfield
  {journal} {\bibinfo  {journal} {Phys. Lett. B}\ }\textbf {\bibinfo {volume}
  {835}},\ \bibinfo {pages} {137581} (\bibinfo {year} {2022})},\ \Eprint
  {http://arxiv.org/abs/2208.13063} {arXiv:2208.13063 [gr-qc]} \BibitemShut
  {NoStop}%
\bibitem [{\citenamefont {Hendi}\ \emph {et~al.}(2023)\citenamefont {Hendi},
  \citenamefont {Hajkhalili},\ and\ \citenamefont {Mahmoudi}}]{Hendi:2022opt}%
  \BibitemOpen
  \bibfield  {author} {\bibinfo {author} {\bibfnamefont {S.~H.}\ \bibnamefont
  {Hendi}}, \bibinfo {author} {\bibfnamefont {S.}~\bibnamefont {Hajkhalili}}, \
  and\ \bibinfo {author} {\bibfnamefont {S.}~\bibnamefont {Mahmoudi}},\
  }\bibfield  {title} {\enquote {\bibinfo {title} {{Thermodynamic stability of
  a new three dimensional regular black hole}},}\ }\href {\doibase
  10.1002/prop.202200101} {\bibfield  {journal} {\bibinfo  {journal}
  {Fortschritte der Physik}\ }\textbf {\bibinfo {volume} {n/a}},\ \bibinfo
  {pages} {2200101} (\bibinfo {year} {2023})},\ \Eprint
  {http://arxiv.org/abs/2204.11558} {arXiv:2204.11558 [gr-qc]} \BibitemShut
  {NoStop}%
\bibitem [{\citenamefont {Boulware}\ and\ \citenamefont
  {Deser}(1985)}]{Boulware:1985wk}%
  \BibitemOpen
  \bibfield  {author} {\bibinfo {author} {\bibfnamefont {David~G.}\
  \bibnamefont {Boulware}}\ and\ \bibinfo {author} {\bibfnamefont {Stanley}\
  \bibnamefont {Deser}},\ }\bibfield  {title} {\enquote {\bibinfo {title}
  {{String Generated Gravity Models}},}\ }\href {\doibase
  10.1103/PhysRevLett.55.2656} {\bibfield  {journal} {\bibinfo  {journal}
  {Phys. Rev. Lett.}\ }\textbf {\bibinfo {volume} {55}},\ \bibinfo {pages}
  {2656} (\bibinfo {year} {1985})}\BibitemShut {NoStop}%
\bibitem [{\citenamefont {Myers}\ and\ \citenamefont
  {Simon}(1988)}]{Myers:1988ze}%
  \BibitemOpen
  \bibfield  {author} {\bibinfo {author} {\bibfnamefont {Robert~C.}\
  \bibnamefont {Myers}}\ and\ \bibinfo {author} {\bibfnamefont {Jonathan~Z.}\
  \bibnamefont {Simon}},\ }\bibfield  {title} {\enquote {\bibinfo {title}
  {{Black Hole Thermodynamics in Lovelock Gravity}},}\ }\href {\doibase
  10.1103/PhysRevD.38.2434} {\bibfield  {journal} {\bibinfo  {journal} {Phys.
  Rev. D}\ }\textbf {\bibinfo {volume} {38}},\ \bibinfo {pages} {2434--2444}
  (\bibinfo {year} {1988})}\BibitemShut {NoStop}%
\bibitem [{\citenamefont {Ma}\ and\ \citenamefont {Zhao}(2014)}]{Ma:2014qma}%
  \BibitemOpen
  \bibfield  {author} {\bibinfo {author} {\bibfnamefont {Meng-Sen}\
  \bibnamefont {Ma}}\ and\ \bibinfo {author} {\bibfnamefont {Ren}\ \bibnamefont
  {Zhao}},\ }\bibfield  {title} {\enquote {\bibinfo {title} {{Corrected form of
  the first law of thermodynamics for regular black holes}},}\ }\href {\doibase
  10.1088/0264-9381/31/24/245014} {\bibfield  {journal} {\bibinfo  {journal}
  {Class. Quant. Grav.}\ }\textbf {\bibinfo {volume} {31}},\ \bibinfo {pages}
  {245014} (\bibinfo {year} {2014})},\ \Eprint {http://arxiv.org/abs/1411.0833}
  {arXiv:1411.0833 [gr-qc]} \BibitemShut {NoStop}%
\bibitem [{\citenamefont {Hull}\ and\ \citenamefont
  {Mann}(2021)}]{Hull:2021bry}%
  \BibitemOpen
  \bibfield  {author} {\bibinfo {author} {\bibfnamefont {Brayden~R.}\
  \bibnamefont {Hull}}\ and\ \bibinfo {author} {\bibfnamefont {Robert~B.}\
  \bibnamefont {Mann}},\ }\bibfield  {title} {\enquote {\bibinfo {title}
  {{Thermodynamics of exotic black holes in Lovelock gravity}},}\ }\href
  {\doibase 10.1103/PhysRevD.104.084032} {\bibfield  {journal} {\bibinfo
  {journal} {Phys. Rev. D}\ }\textbf {\bibinfo {volume} {104}},\ \bibinfo
  {pages} {084032} (\bibinfo {year} {2021})},\ \Eprint
  {http://arxiv.org/abs/2102.05282} {arXiv:2102.05282 [gr-qc]} \BibitemShut
  {NoStop}%
\bibitem [{\citenamefont {Mora}\ \emph {et~al.}(2004)\citenamefont {Mora},
  \citenamefont {Olea}, \citenamefont {Troncoso},\ and\ \citenamefont
  {Zanelli}}]{Mora:2004rx}%
  \BibitemOpen
  \bibfield  {author} {\bibinfo {author} {\bibfnamefont {P.}~\bibnamefont
  {Mora}}, \bibinfo {author} {\bibfnamefont {R.}~\bibnamefont {Olea}}, \bibinfo
  {author} {\bibfnamefont {R.}~\bibnamefont {Troncoso}}, \ and\ \bibinfo
  {author} {\bibfnamefont {J.}~\bibnamefont {Zanelli}},\ }\bibfield  {title}
  {\enquote {\bibinfo {title} {{Vacuum energy in odd-dimensional AdS
  gravity}},}\ }\href@noop {} {\  (\bibinfo {year} {2004})},\ \Eprint
  {http://arxiv.org/abs/hep-th/0412046} {arXiv:hep-th/0412046} \BibitemShut
  {NoStop}%
\bibitem [{\citenamefont {Christiansen}\ \emph {et~al.}(2023)\citenamefont
  {Christiansen}, \citenamefont {Estrada}, \citenamefont {Cunha}, \citenamefont
  {Furtado},\ and\ \citenamefont {Muniz}}]{Christiansen:2022ebo}%
  \BibitemOpen
  \bibfield  {author} {\bibinfo {author} {\bibfnamefont {H.~R.}\ \bibnamefont
  {Christiansen}}, \bibinfo {author} {\bibfnamefont {Milko}\ \bibnamefont
  {Estrada}}, \bibinfo {author} {\bibfnamefont {M.~S.}\ \bibnamefont {Cunha}},
  \bibinfo {author} {\bibfnamefont {J.}~\bibnamefont {Furtado}}, \ and\
  \bibinfo {author} {\bibfnamefont {C.~R.}\ \bibnamefont {Muniz}},\ }\bibfield
  {title} {\enquote {\bibinfo {title} {{New regular 2+1 black hole solutions
  from bilocal gravity}},}\ }\href {\doibase 10.1142/S0218271823500414}
  {\bibfield  {journal} {\bibinfo  {journal} {International Journal of Modern
  Physics D}\ } (\bibinfo {year} {2023}),\ 10.1142/S0218271823500414},\ \Eprint
  {http://arxiv.org/abs/2201.12814} {arXiv:2201.12814 [gr-qc]} \BibitemShut
  {NoStop}%
\bibitem [{\citenamefont {Miskovic}\ and\ \citenamefont
  {Olea}(2007)}]{Miskovic:2007mg}%
  \BibitemOpen
  \bibfield  {author} {\bibinfo {author} {\bibfnamefont {Olivera}\ \bibnamefont
  {Miskovic}}\ and\ \bibinfo {author} {\bibfnamefont {Rodrigo}\ \bibnamefont
  {Olea}},\ }\bibfield  {title} {\enquote {\bibinfo {title} {{Counterterms in
  Dimensionally Continued AdS Gravity}},}\ }\href {\doibase
  10.1088/1126-6708/2007/10/028} {\bibfield  {journal} {\bibinfo  {journal}
  {JHEP}\ }\textbf {\bibinfo {volume} {10}},\ \bibinfo {pages} {028} (\bibinfo
  {year} {2007})},\ \Eprint {http://arxiv.org/abs/0706.4460} {arXiv:0706.4460
  [hep-th]} \BibitemShut {NoStop}%
\bibitem [{\citenamefont {Adler}\ \emph {et~al.}(2001)\citenamefont {Adler},
  \citenamefont {Chen},\ and\ \citenamefont {Santiago}}]{Adler:2001vs}%
  \BibitemOpen
  \bibfield  {author} {\bibinfo {author} {\bibfnamefont {Ronald~J.}\
  \bibnamefont {Adler}}, \bibinfo {author} {\bibfnamefont {Pisin}\ \bibnamefont
  {Chen}}, \ and\ \bibinfo {author} {\bibfnamefont {David~I.}\ \bibnamefont
  {Santiago}},\ }\bibfield  {title} {\enquote {\bibinfo {title} {{The
  Generalized uncertainty principle and black hole remnants}},}\ }\href
  {\doibase 10.1023/A:1015281430411} {\bibfield  {journal} {\bibinfo  {journal}
  {Gen. Rel. Grav.}\ }\textbf {\bibinfo {volume} {33}},\ \bibinfo {pages}
  {2101--2108} (\bibinfo {year} {2001})},\ \Eprint
  {http://arxiv.org/abs/gr-qc/0106080} {arXiv:gr-qc/0106080} \BibitemShut
  {NoStop}%
\bibitem [{\citenamefont {Dymnikova}(1992)}]{Dymnikova:1992ux}%
  \BibitemOpen
  \bibfield  {author} {\bibinfo {author} {\bibfnamefont {I.}~\bibnamefont
  {Dymnikova}},\ }\bibfield  {title} {\enquote {\bibinfo {title} {{Vacuum
  nonsingular black hole}},}\ }\href {\doibase 10.1007/BF00760226} {\bibfield
  {journal} {\bibinfo  {journal} {Gen. Rel. Grav.}\ }\textbf {\bibinfo {volume}
  {24}},\ \bibinfo {pages} {235--242} (\bibinfo {year} {1992})}\BibitemShut
  {NoStop}%
\bibitem [{\citenamefont {Estrada}\ and\ \citenamefont
  {Aros}(2019)}]{Estrada:2019qsu}%
  \BibitemOpen
  \bibfield  {author} {\bibinfo {author} {\bibfnamefont {Milko}\ \bibnamefont
  {Estrada}}\ and\ \bibinfo {author} {\bibfnamefont {Rodrigo}\ \bibnamefont
  {Aros}},\ }\bibfield  {title} {\enquote {\bibinfo {title} {{Regular black
  holes with $\Lambda>0$ and its evolution in Lovelock gravity}},}\ }\href
  {\doibase 10.1140/epjc/s10052-019-7316-0} {\bibfield  {journal} {\bibinfo
  {journal} {Eur. Phys. J. C}\ }\textbf {\bibinfo {volume} {79}},\ \bibinfo
  {pages} {810} (\bibinfo {year} {2019})},\ \Eprint
  {http://arxiv.org/abs/1906.01152} {arXiv:1906.01152 [gr-qc]} \BibitemShut
  {NoStop}%
\bibitem [{\citenamefont {Kuntz}\ and\ \citenamefont
  {Da~Rocha}(2020)}]{Kuntz:2019gka}%
  \BibitemOpen
  \bibfield  {author} {\bibinfo {author} {\bibfnamefont {Iber\^e}\ \bibnamefont
  {Kuntz}}\ and\ \bibinfo {author} {\bibfnamefont {Rold\~ao}\ \bibnamefont
  {Da~Rocha}},\ }\bibfield  {title} {\enquote {\bibinfo {title} {{GUP black
  hole remnants in quadratic gravity}},}\ }\href {\doibase
  10.1140/epjc/s10052-020-8049-9} {\bibfield  {journal} {\bibinfo  {journal}
  {Eur. Phys. J. C}\ }\textbf {\bibinfo {volume} {80}},\ \bibinfo {pages} {478}
  (\bibinfo {year} {2020})},\ \Eprint {http://arxiv.org/abs/1909.05552}
  {arXiv:1909.05552 [hep-th]} \BibitemShut {NoStop}%
\bibitem [{\citenamefont {Ward}(2006)}]{Ward:2006hf}%
  \BibitemOpen
  \bibfield  {author} {\bibinfo {author} {\bibfnamefont {B.~F.~L.}\
  \bibnamefont {Ward}},\ }\bibfield  {title} {\enquote {\bibinfo {title}
  {{Resummed quantum gravity}},}\ }\href {\doibase 10.1142/S021827180801236X}
  {\bibfield  {journal} {\bibinfo  {journal} {Conf. Proc. C}\ }\textbf
  {\bibinfo {volume} {060726}},\ \bibinfo {pages} {1233--1237} (\bibinfo {year}
  {2006})},\ \Eprint {http://arxiv.org/abs/hep-ph/0610232}
  {arXiv:hep-ph/0610232} \BibitemShut {NoStop}%
\bibitem [{\citenamefont {Zhang}\ \emph {et~al.}(2020)\citenamefont {Zhang},
  \citenamefont {Ma}, \citenamefont {Song},\ and\ \citenamefont
  {Zhang}}]{Zhang:2020qxw}%
  \BibitemOpen
  \bibfield  {author} {\bibinfo {author} {\bibfnamefont {Cong}\ \bibnamefont
  {Zhang}}, \bibinfo {author} {\bibfnamefont {Yongge}\ \bibnamefont {Ma}},
  \bibinfo {author} {\bibfnamefont {Shupeng}\ \bibnamefont {Song}}, \ and\
  \bibinfo {author} {\bibfnamefont {Xiangdong}\ \bibnamefont {Zhang}},\
  }\bibfield  {title} {\enquote {\bibinfo {title} {{Loop quantum Schwarzschild
  interior and black hole remnant}},}\ }\href {\doibase
  10.1103/PhysRevD.102.041502} {\bibfield  {journal} {\bibinfo  {journal}
  {Phys. Rev. D}\ }\textbf {\bibinfo {volume} {102}},\ \bibinfo {pages}
  {041502} (\bibinfo {year} {2020})},\ \Eprint
  {http://arxiv.org/abs/2006.08313} {arXiv:2006.08313 [gr-qc]} \BibitemShut
  {NoStop}%
\bibitem [{\citenamefont {Ali}\ and\ \citenamefont
  {Khalil}(2016)}]{Ali:2015tva}%
  \BibitemOpen
  \bibfield  {author} {\bibinfo {author} {\bibfnamefont {Ahmed~Farag}\
  \bibnamefont {Ali}}\ and\ \bibinfo {author} {\bibfnamefont {Mohammed~M.}\
  \bibnamefont {Khalil}},\ }\bibfield  {title} {\enquote {\bibinfo {title}
  {{Black Hole with Quantum Potential}},}\ }\href {\doibase
  10.1016/j.nuclphysb.2016.05.005} {\bibfield  {journal} {\bibinfo  {journal}
  {Nucl. Phys. B}\ }\textbf {\bibinfo {volume} {909}},\ \bibinfo {pages}
  {173--185} (\bibinfo {year} {2016})},\ \Eprint
  {http://arxiv.org/abs/1509.02495} {arXiv:1509.02495 [gr-qc]} \BibitemShut
  {NoStop}%
\end{thebibliography}%

\end{document}